# Twist-Controlled Wire Metasurfaces

Ingrid Torres[1], and Alex Krasnok[1,2*]

[1]Department of Electrical and Computer Engineering, Florida International University, Miami, Florida 33174, USA

[2]Knight Foundation School of Computing and Information Sciences, Florida International University, Miami, FL 33199, USA

*To whom correspondence should be addressed: akrasnok@fiu.edu

## Abstract

Twistronics, originally conceptualized within the electronics domain to modulate electronic properties through the twist angle between stacked two-dimensional (2D) materials, presents a groundbreaking approach in material science. This concept's extension to photonics, especially using metasurfaces, offers a promising avenue for manipulating light at subwavelength scales across a broad electromagnetic spectrum. Nevertheless, the possibilities that twistronics presents within the realm of photonics are still significantly untapped. In this work, we explore a photonic configuration consisting of a dual-layer wire metasurface operating within the microwave frequency spectrum, where the interlayer twist angle governs the system. Our findings reveal that such manipulation significantly alters the electromagnetic response of the structure, leading to enhanced resonances, tunability, and mode coupling. This twist-induced modulation results in a resonance shift towards the low-frequency range, effectively miniaturizing the structure's electrical dimension. Our study demonstrates the feasibility of integrating twistronics into photonic designs and opens new pathways for developing tunable and reconfigurable photonic devices.

## I.    Introduction

The advent of two-dimensional (2D) materials, initiated by the groundbreaking discovery of graphene[1–3], has opened a new frontier in the exploration of materials science and condensed matter physics. These materials exhibit a plethora of unique properties—from electronic and optical to thermal and mechanical characteristics—that make them ideal for developing next-generation ultrathin nanodevices[4–6]. A significant advancement in this domain is the creation of 2D material heterostructures, achieved by stacking layers of these materials with atomic precision at specific twist angles[7–10]. This methodology has unveiled a host of intriguing phenomena, such as the emergence of superconductivity in twisted bilayer graphene[8] and interlayer magnetism in stacked 2D magnetic materials[11], among others. These discoveries have given rise to the field of twistronics and moiré structures, which focuses on manipulating the electron wavefunction through rotational adjustments[12–16], offering profound implications for material science.

Concurrently, the field of photonics has seen metasurfaces—flat, quasi-periodic structures with subwavelength periodicity—emerge as a key research area, showcasing the unmatched ability to manipulate electromagnetic waves at scales smaller than their wavelength[17–21]. The flexibility in designing meta-atoms[22], in terms of their shape, size, orientation, and other parameters, allows for comprehensive control over the electromagnetic field's phase, amplitude, and polarization. This attribute opens doors to a myriad of applications, such as enhanced solar cells, wireless communications, holographic displays, and super-resolution lenses[23–26]. Despite the profound impact of twistronics in electronic systems, its application to photonics using metasurfaces remains relatively uncharted[27,28]. This gap presents a compelling research opportunity, as metasurfaces offer a versatile platform for the extreme manipulation of light across various parts of the electromagnetic spectrum, from radio frequency to visible frequencies. While the concept of photonic twistronics—particularly through the stacking of metasurfaces—presents a promising avenue, it has not been extensively explored. However, realizing these potentials requires intricate manipulation of photonic structures at the subwavelength level, akin to the precise rotational control in twistronic heterostructures.

In this work, we delve into the twistronic effects within a photonic framework, utilizing a double-layer wire metasurface (DLWM) structure in the microwave frequency range. Wire metamaterials and their 2D counterparts, wire metasurfaces (WMs), with their dense arrays of metal wires embedded in dielectric matrices, span a broad spectrum from radio frequency to visible light[29,30]. Their optical properties, adjustable through wire size and arrangement, provide unique electromagnetic features, facilitating advancements in photonics such as subwavelength imaging[31]. Through theoretical analysis and experimental validation, we demonstrate that the DLWM structure exhibits strong resonances characterized by intense electric and magnetic fields, alongside significant tunability and strong mode coupling influenced by the relative rotation of the layers. Notably, rotational manipulation shifts the resonances towards the low-frequency range, effectively miniaturizing the structure's electrical dimensions. Our findings contribute to the advancement of photonic platforms by integrating twistronics into the design of tunable and reconfigurable structures and highlight the potential for a wide array of applications. For example, WMs are critical for enhancing MRI technology by improving signal-to-noise ratios and resolution[31,32]. Therefore, the tunability of these structures holds fundamental interest and practical significance for real-world applications.

## II. Results and discussion

The geometry of the DLWM structure under investigation is schematically depicted in **Figure 1**, consisting of two identical superimposed wire metasurfaces. Each metasurface is based on a resonant array of metallic wires, specifically 31 aluminum (Al) wires with a length $L = 350$ mm, a radius of 0.9 mm, and separated by approximately $a = 5$ mm, immersed in free space. In the experiment, these WMs are separated by a foam slab with a permittivity similar to air, verified to minimally affect the measurement results. This configuration facilitated the separation and rotation of the WMs. Given the wire length of $L = 350$ mm, the fundamental dipole resonance of each wire, occurring at $\lambda = 2 \cdot L$ where $\lambda$ is the wavelength, is expected at a frequency $f = 428$ MHz. This resonance mode, known as the first Fabry–Pérot mode[30,33], signifies a standing wave of current in each wire with a maximum at the wire's center (and corresponding voltage maxima at the wire ends), leading to a strongly enhanced magnetic field at the center of the wire with a suppressed electric field (and, conversely, an enhanced electric field at the wire ends).

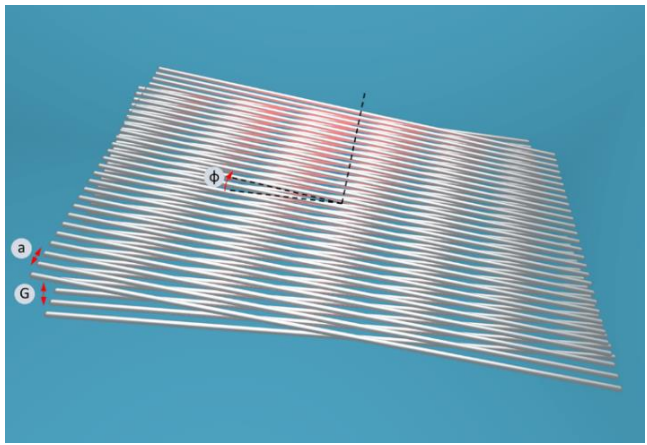

**Figure 1:** Schematic representation of the twist-controlled double-layer wire metasurface (DLWM) structure.

We conducted rigorous full-wave numerical simulations using CST Microwave Studio[34] to gain insights into the structure, exciting the DLWM structure with a linearly polarized plane wave between 300-500 MHz. The wave's electric field (E-field), with a strength of 1 V/m, was oriented along the wires, and the magnetic field (H-field), with a strength of 1/377 A/m, was orthogonal to the DLWM plane. The results in **Figure 2** show that without rotation (0°), the structure exhibits a band of eigenmodes centered around 420 MHz, aligning well with the fundamental mode estimation above. This band widens due to interwire couplings and contains peaks equal to the number of wires in the structure. We explored the tunability effect upon the relative rotation of the WMs. One of the WMs was rotated counterclockwise from 0° to 50° in 10° increments. Both the H-field and E-field of the structure began to change spectrally and spatially as the layer rotated. **Figure 2(a)** displays the spectra of the H-field in the center of the DLWM, placed 2 mm above one of the arrays for various relative rotation angles, with a separation gap (G) of 2 mm between the layers. All results in **Figure 2** are normalized to the field strength of the plane wave. Upon rotation, many resonances appear in the lower frequency range, absent at a 0° angle (parallel WMs' configuration). The position, shape, and number of these resonances strongly depend on the rotation angle, allowing for fine-tuning of the resonant properties of the structure. In our simulations, we utilized realistic material parameters for pure Al. To ensure convergence, we also introduced a minor loss to the adjacent dielectric material, characterized by a permittivity of 1 + i0.001.

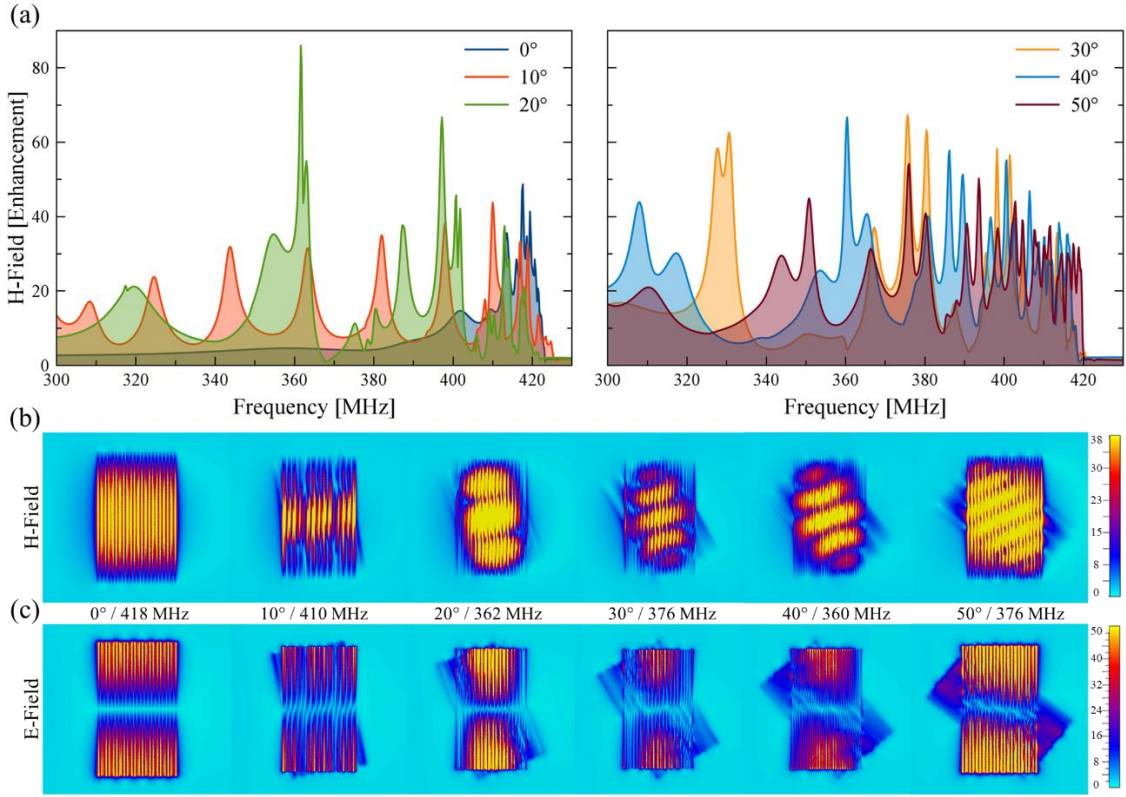

**Figure 2:** Tunable resonant behavior and field distribution of DLWM structure under rotation. (a) H-field spectra in the center of the DLWM, excited by a linearly polarized plane wave between 300-500 MHz, illustrating eigenmode bands around 420 MHz at 0° and the emergence of numerous resonances at lower frequencies as one WM is rotated from 0° to 50° in 10° increments. Results are normalized to the incident plane wave's field strength, showcasing how rotational adjustments fine-tune the resonant properties due to changes in interwire coupling. (b) H-field and (c) E-field enhancement spatial distribution maps for various rotation angles (0, 10, 20, 30, 40, 50 degrees) at major resonant frequencies.

**Figure 2(b)** shows the distributions of the H-field enhancement, and **Figure 2(c)** shows the distributions of the E-field enhancement at major resonances for different rotation angles (0, 10, 20, 30, 40, 50 degrees). At a 0° angle, the maximum H-field of 49 at 418 MHz covers a large, centralized area of the structure, while the maximum E-field is concentrated towards the wire edges, and both show nonuniform distribution—a response consistent with previous works[30,32]. As the angle increases, changes in the distribution patterns of both fields are observed, with significant variations at 20° and 30°, leading to the localization of the E-field at the wire edges and the H-field localized in the central area with the formation of the field distribution periodic pattern, inherited from the moiré pattern (Figure 1). Notably, at 20°, the H-field enhancement reaches 86 at the frequency of 362 MHz. The phenomena observed intensify up to a 40° rotation, with notable adjustments in the field's intensity and distribution areas. At 50°, the E-field extends towards the center of the structure, while the maximum H-field decreases to 54, increasing the overall H-field area but concentrating in five diagonal areas, mirroring the rotation angle in a clockwise direction. These results illustrate the high dependency of the magnetic and electric fields, and their distribution areas, on the rotation angle of the metasurface structure, highlighting the capability for targeted manipulation of the field distributions within specific, localized areas of interest.

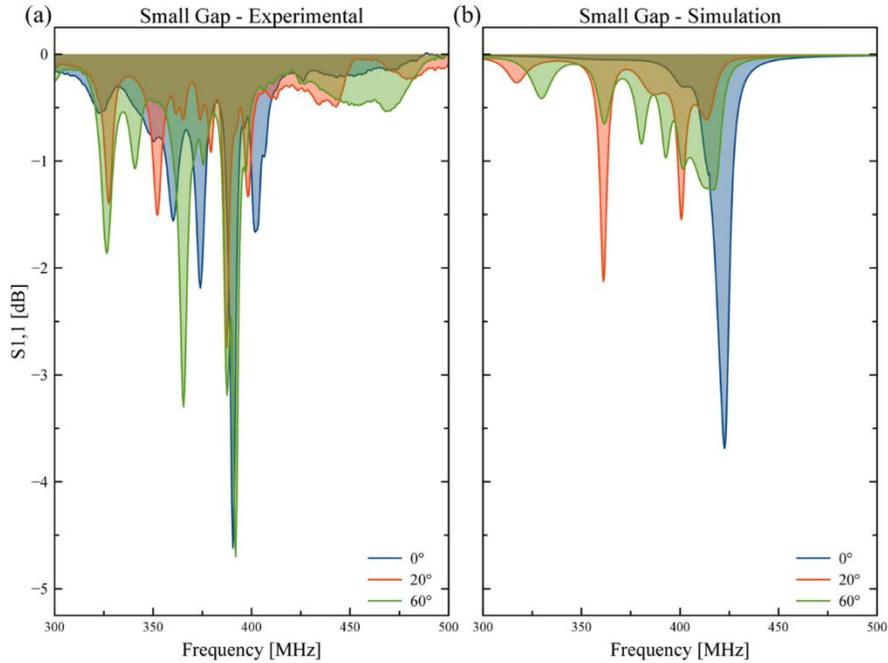

**Figure 3:** Experimental and simulated reflection coefficient ($S_{11}$) analysis of the DLWM structure. (a) Measured $S_{11}$ values for a small gap (G = 5mm) at 0°, 20°, and 60° rotations, illustrating the main resonance frequency around 390 MHz and its subsequent splitting and shift towards lower frequencies with rotation. (b) Full-wave simulation results for a small gap (G = 2mm) at the same rotational condition.

Next, we advanced to the experimental investigation of the fabricated structures. In the experimental setup, the two WMs were aligned in parallel and separated by a thin foam board with permittivity similar to air. The S-parameters were measured using a Vector Network Analyzer (VNA), Agilent 8720ES/HP 8720ES. A small loop antenna with a 10 mm radius, connected to the VNA via a coaxial cable and positioned 2 mm above one of the WM, was used as a field probe. The DLWM structure was tested for two different gap sizes: small (G=5 mm) and large (G=30 mm).

The measurement results for the small gap (G = 5mm) are shown in **Figure 3(a)**, with CST full-wave simulation results for comparison in **Figure 3(b)**. With the two WMs parallel (0°), the main resonance frequency is around 391 MHz. The WM further from the loop was then gradually rotated counterclockwise in 10° increments, causing the main resonance frequency to split and shift toward the lower end of the frequency spectrum. At 20° and 60°, the DLWM structure demonstrates the splitting of the original resonance into several bands, with the main resonance slightly shifting. The bands become more pronounced at 60°. The DLWM structure was simulated to verify these effects with a small gap between the layer arrays and a loop antenna positioned 2 mm above one of the WM. The WM farther from the loop was rotated, and the reflection coefficient $S_{11}$ was calculated, **Figure 3(b)**. The simulation results confirm that the original resonance frequency splits, and the main resonance shifts towards the lower frequency range as the rotation increases, aligning with the experimental findings.

As the gap between the arrays widens to 30mm, the tuning effect diminishes, with only minor changes observed upon rotation, as depicted in **Figure 4**. Experimentally, when one of the array layers rotates, the original resonance frequency remains largely unchanged, with slight splitting observed between 315 – 410 MHz, while the main resonance frequency stays around 394 MHz. Similarly, simulation results indicate that the original resonance frequency remains steady, fluctuating slightly between 421 – 423 MHz, and exhibits subtle splitting between 325 – 425 MHz. This phenomenon can be attributed to the reduced coupling effect between the WMs as the gap between them increases. Although the arrays do not become completely decoupled, some splitting suggests a weakened yet persistent interaction.

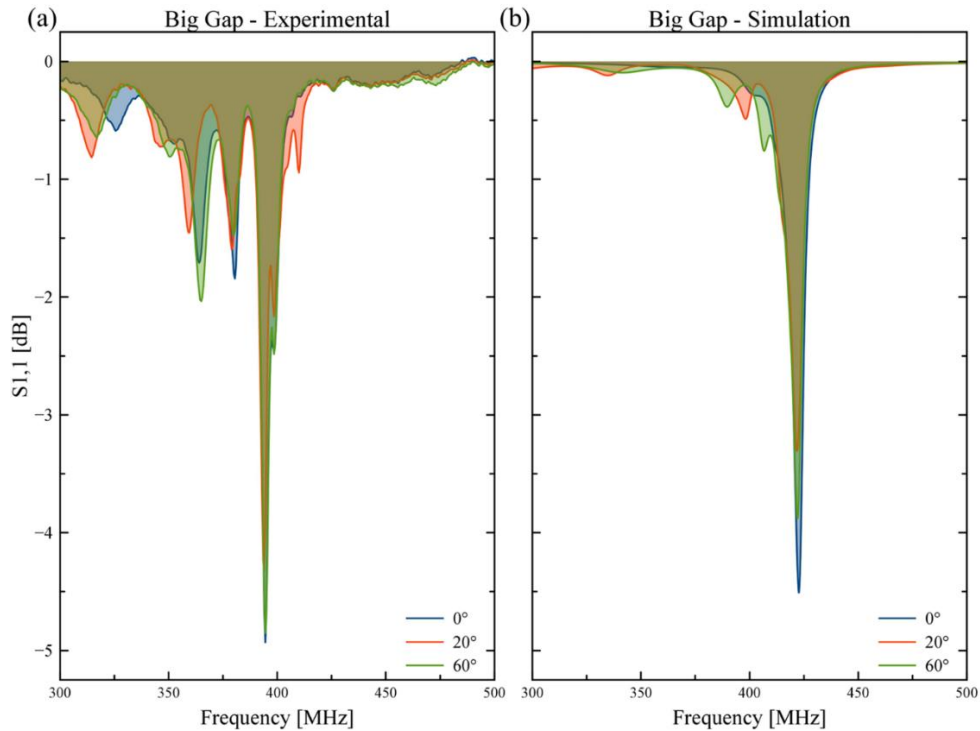

**Figure 4:** Reflection coefficient ($S_{11}$) analysis for the DLWM structure with a big gap. (a) Experimental measurements showing the original resonance frequency's minor change upon relative rotations for a big gap (G = ~30 mm). (b) Simulated results for a big gap (G = 10 mm) confirming the experimental findings.

Finally, the experimental frequency range was expanded to 200-500 MHz to explore the full extent of the tuning effect, as illustrated in **Figure 5**. The original resonance frequency splits into several bands, showing significant tuning across different frequencies as the rotation degree changes. At a 0° rotation angle, with both array layers parallel, the resonance frequency is observed at 409 MHz. Note that the position of the loop antenna probe was changed in this experiment compared to Figure 3 to improve the coupling to the lowest frequency mode. With a gradual angle change from 0 to 20°, the main resonance frequency shifts to 285 MHz. Further rotation leads to additional shifts and the emergence of multiple resonances. At 40°, resonances are seen at 250 MHz and 234 MHz. The resonance with the lowest frequency is observed at 60° at 210 MHz. This capacity for tuning through rotational adjustments makes the metasurface structure appealing for various practical applications. Notably, the lowest observed resonant frequency of 210 MHz corresponds to a wavelength of 1427 mm, rendering the DLWM structure, with a size of L=350 mm, four times smaller than the wavelength, thus making the system electrically small[35]. This feature is particularly relevant for applications in antennas and sensors.

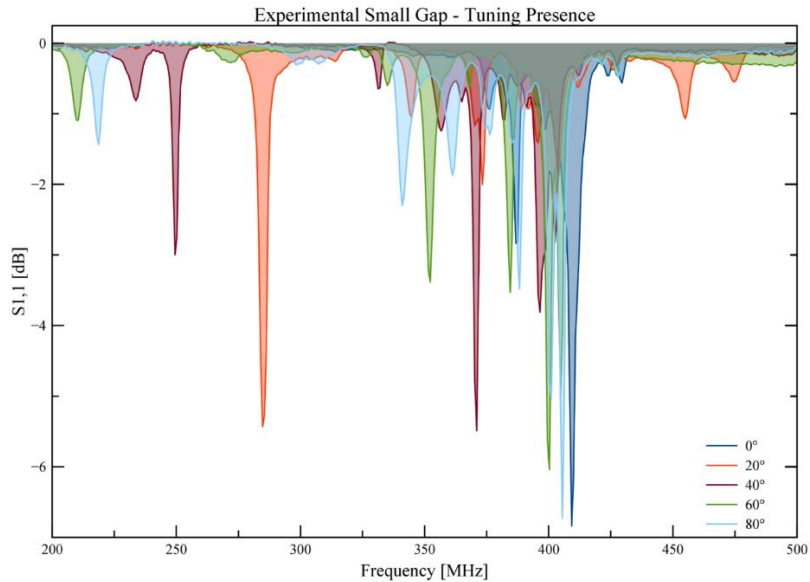

**Figure 5:** Experimental reflection coefficient ($S_{11}$) of the DLWM structure with expanded frequency range (200-500 MHz). This figure illustrates the tuning effect through the splitting of the original resonance frequency at rotation angles of 0°, 20°, 40°, and 60°. It highlights the gradual shift and splitting of the main resonance frequency from 409 MHz at 0° to 210 MHz at 60°, demonstrating the structure's capability for significant frequency tuning via rotation.

### III.    Conclusions

In conclusion, our investigation unveiled an approach for modulating the electromagnetic characteristics of wire metasurfaces by manipulating the twist angle between layers in a bi-layer arrangement operating at microwave frequencies. This method notably enhanced resonance behaviors, tunability, and mode coupling, facilitating a resonance frequency shift towards the lower spectrum. This shift enabled a diminution of the metasurface's electrical footprint. Our findings underscore the potential of twist angle control as a powerful tool for designing and developing tunable and reconfigurable photonic devices, heralding a novel technique for meticulously manipulating electromagnetic waves on subwavelength dimensions. The implications of this research are far-reaching, offering substantial prospects for its application in enhancing MRI technologies, bolstering communication systems, and refining sensing capabilities. This work contributes to the broader understanding of photonic structure manipulation and sets the stage for future explorations into the versatile applications of twist-controlled metasurfaces in advanced technological domains.

**Acknowledgments**

The authors thank the ECE department of Florida International University.

**AUTHOR DECLARATIONS**

**Conflict of Interest**

The authors have no conflicts to disclose.

**Author Contributions**

Ingrid Torres and Alex Krasnok collaboratively contributed to experiments, analysis, investigation, and methodology. Torres led numerical simulations and drafted the manuscript, while Krasnok managed the project, provided supervision, and enhanced the manuscript through critical review and editing.

**DATA AVAILABILITY**

The data that support the findings of this study are available from the corresponding authors upon reasonable request.